# Complex magnetic properties in the mixed 4f -5d double perovskite iridates Ln$_2$ZnIrO$_6$ (Ln = Nd, Sm, Eu & Gd)

M. Vogl[1],* R. Morrow[1], A.A. Aczel[3,4], R. Beltran Rodriguez[1], A.U.B. Wolter[1], S. Wurmehl[1,2], S. Aswartham[1],# and B. Büchner[1,2]

[1]Institute for Solid State Research, Leibniz IFW Dresden, Helmholtzstr. 20, 01069 Dresden, Germany

[2]Institut für Festkörperphysik, TU Dresden, D-01062 Dresden, Germany

[3]Neutron Scattering Division, Oak Ridge National Laboratory, Oak Ridge, Tennessee 37831, USA and

[4]Department of Physics and Astronomy, University of Tennessee, Knoxville, Tennessee 37996, USA

(Dated: October 29, 2019)

## Abstract

In this work, we report on the synthesis and magnetic properties of a series of double perovskites Ln$_2$ZnIrO$_6$ with Ln = Nd, Sm, Eu & Gd. These compounds present new examples of the rare case of double perovskites (general formula A$_2$BB'O$_6$) with a magnetic 4*f*-ion on the *A*-site in combination with the strongly spin-orbit coupled 5*d*-transition metal ion Ir$^{4+}$ on the *B*-sublattice. We discuss the impact of different rare earths on the macroscopic magnetic properties. Gd$_2$ZnIrO$_6$ and Eu$_2$ZnIrO$_6$ show weak canted antiferromagnetic order below $T_N$ = 23 K and $T_N$ = 12 K, respectively. Sm$_2$ZnIrO$_6$ orders antiferromagnetically at $T_N$ = 13 K. Nd$_2$ZnIrO$_6$ exhibits complex magnetic properties with strong field dependence ranging from a two-step behavior at H = 0.01 T to an antiferromagnetic ground state at intermediate external fields and a spin-flop phase at H ≥ 4 T, which suggests complex interplay between Nd$^{3+}$ and Ir$^{4+}$. To further shed light on this magnetic interaction, the magnetic structure of Nd$_2$ZnIrO$_6$'s ground state is examined via neutron powder diffraction.

*m.vogl@ifw-dresden.de
#s.aswartham@ifw-dresden.de

## I. INTRODUCTION

Oxides of 5d-metals have been the subject of intense studies over the past years. The unique interplay of crystal field splitting, spin-orbit coupling (SOC) and Coulomb repulsion, leading to novel correlated ground states, have made these materials an intriguing field to study fundamental physics [1-3]. One common motif in which 5*d*-oxides can be synthesized is the double perovskite structure. Double perovskites with the general formula $A_2BB'O_6$ offer the opportunity to tune and examine their ground states by inserting a variety of different ions on both the *A*- and *B*-sites [4-6]. This has been widely used in the past to study the interplay of 3*d*- and 5*d*-metals on either *B*-site, in compounds such as $La_2TMIrO_6$ (TM = Co, Ni, Cu) and $AE_2TMOsO_6$ (AE = Ca, Sr; TM = Cr, Fe, Co, Ni) [7-14]. The physical properties of double perovskites with two magnetic *B*-cations have been thoroughly studied for both iridates and osmates. The case of double perovskite materials with a magnetic 4*f*-ion on the *A*-site in combination with a 5*d*-element as the only magnetic *B*-cation however is mainly known for osmates, such as $Ln_2NaOsO_6$ and $Ln_2LiOsO_6$ [15; 16; 33]. For the iridates, the only known study treats $Ln_2MgIrO_6$ [17]. In that work, there is no clear trend of the magnetic properties as a function of the size of the lanthanide-ion. In addition, a comparison of $Ln_2ZnIrO_6$ and $Ln_2MgIrO_6$ may shed light on the effect of the non-magnetic *B*-cation on the magnetic properties. The different magnetic properties of the antiferromagnet $La_2MgIrO_6$ and the canted antiferromagnet $La_2ZnIrO_6$ suggest a significant influence [9; 19-22]. This phenomenon has been theoretically discussed in osmates [18], but remains an open question for the iridate case. For this study, a new series of polycrystalline double perovskites, $Ln_2ZnIrO_6$ (Ln = Gd, Eu, Sm & Nd), were synthesized, carefully characterized and their magnetic properties were investigated.

## II. EXPERIMENTAL DETAILS

Polycrystalline samples of $Ln_2ZnIrO_6$ (Ln = Nd, Sm, Eu & Gd) were synthesized from stoichiometric amounts of the corresponding lanthanide oxide $Ln_2O_3$, ZnO and $IrO_2$. Prior to this reaction $Ln_2O_3$ was heated to $900^0C$ for 12 h in order to remove moisture. The reaction mixture was heated to $1100^0C$ with a heating rate of $200^0C/h$ and held there for 30 h. After cooling to room temperature with a rate of $150^0C/h$ a regrinding step followed. In a second annealing step, the reaction mixture was heated to $1200\ ^0C$ with a rate of $200\ ^0C/h$ and kept at that temperature for 50 h. The cooling rate of the 2nd heating step was $50^0C/h$ down to $500^0C$ followed by switching off the furnace. For structural analysis, X-ray diffraction (XRD) measurements were carried out in the transmission method on a StoeStadi-Powder diffractometer with Co-$K_{\alpha 1}$ radiation. Neutron powder diffraction (NPD) data were collected at the High Flux Isotope Reactor

(HFIR) facility at ORNL on the powder diffractometer HB-2A using a 5 grams of polycrystalline sample of $Nd_2ZnIrO_6$ loaded in a 5 mm diameter V can to minimize Ir neutron absorption. Data were collected at 4 K, 15 K and 30 K with a neutron wavelength of 2.41 Å, while further data were collected at 4 K with a neutron wavelength of 1.54 Å. Additional NPD data were collected on the triple-axis spectrometer HB-1A at HFIR using the same sample now loaded in an annular Al can with a 1 mm annulus. Data were collected at a variety of temperatures using a neutron wavelength of 2.36 Å to measure the order parameters of a subset of the magnetic Bragg peaks. In the analysis of the HB-2A NPD diffractograms, Al and Cu contributions from the sample environment were considered. Rietveld refinement of the atomic structures was performed using the Full-Prof software package [23; 24]. For the refinement of the magnetic structure of $Nd_2ZnIrO_6$, the software SARAh was used in addition to FullProf [25; 26]. The resulting magnetic structure has been visualized with the program VESTA [27]. Scanning electron microscopy (SEM) and energy dispersive x-ray-spectroscopy (EDXS) were performed on a nanoSEM by FEI. The analysis was carried out on pressed powder pellets. Magnetic measurements were performed using a Quantum Design MPMS-XL SQUID-magnetometer. Temperature-dependent magnetization was measured in both field-cooled and zero-field cooled mode. Curie-Weiss analysis was performed after subtracting the temperature-independent portion of each sample's magnetization. Specific heat measurements were performed on a pressed powder sample between 1.8 - 30 K using a heat-pulse relaxation method in a Physical Properties Measurement System (PPMS) from Quantum Design. The heat capacity of the sample holder (addenda) was determined prior to the measurements in order to separate the heat capacity contribution of the sample from the total heat capacity.

III. RESULTS

A. Crystal structure and composition

The phase purity and crystal structure of all samples was probed by powder X-ray diffraction. Figure 1 shows a comparison of the XRD pattern of all compounds of the series. There are no significant changes in the pattern as a function of the *A*-cation radius. The Goldschmidt tolerance factor can be used as a tool to predict the structure of perovskite type materials [28]. For $Ln_2ZnIrO_6$, the values fall in the range between 0.85 (Ln = Gd) and 0.87 (Ln = Nd). These values are significantly below t = 1 and are usually found for monoclinic perovskites. Rietveld refinement confirms that all compounds of the series adopt a monoclinic double perovskite structure with the space group *P 21/n* (Sp. gr. 14). The same space group has been reported for similar compounds like $La_2ZnIrO_6$ and $Ln_2MgIrO_6$ [9; 17; 19; 20; 37]. The pattern for Ln = Eu, Sm & Nd show no signs of a secondary phase. For Ln=Gd about 3% of remaining $Gd_2O_3$ is found. The lattice parameters derived from the structure refinement are shown in Fig.2. A linear dependence of a and c on the radius of the A-cation is observed, while b stays approximately constant [29]. The chemical compositions of all samples were examined by SEM and EDXS. Topology and chemical contrast images taken by SEM confirm chemical homogeneity. Figure 3 shows the SEM images of $Eu_2ZnIrO_6$ as a representative example. Compositions close to the nominal ones were found for all samples via EDXS (table 1).

**B. Magnetic properties**

**a) $Gd_2ZnIrO_6$**
The magnetization data of $Gd_2ZnIrO_6$, the compound with the smallest rare earth ion, shows a ferromagnetic like (FM) transition at 24 K (Fig. 4a). The transition temperature of $Gd_2ZnIrO_6$, and all other compounds in this report that show a FM-like transition, were determined on the data measured in an external field of 1 T via the 2$^{nd}$ derivative method, i.e. the transition temperature is defined as the inflection point of the magnetization curve. At low fields (H ≤ 0.1 T) a splitting of the FC- and ZFC-curves is observed at 15 K. This is a common feature for ferromagnetic compounds, caused by domain formation. The field-dependent magnetization at 5 K confirms the FM-ground state (Fig. 4b). A small hysteresis is observed with a remanent magnetization of 1.5 $\mu_B$/f.u. and a coercive field of 135 Oe. Notably, the magnetic moment keeps increasing up to an applied field of 7 T. No saturation is achieved. This may be a sign for a ferrimagentic alignment of $Gd^{3+}$- and $Ir^{4+}$-spins or a canted spin structure. Curie-Weiss analysis of the inverse magnetization yields a Curie-Weiss temperature $\theta_{CW}$ = 2.0 K and an effective magnetic moment $\mu_{eff}$ = 11.3 $\mu_B$/f.u. (Fig. 4c). The positive value of CW signals net-ferromagnetic interactions in the compound, in agreement with the ferromagnetic transition found in the M(T) data. The theoretically expected value of $\mu_{eff}$ can be calculated by taking the square root of the sum of squares of the moments of all magnetic ions in the unit cell. In the case of $Gd_2ZnIrO_6$ it amounts to 11.36 $\mu_B$/f.u., in good agreement with the experimentally derived value.

**b) $Eu_2ZnIrO_6$**

In the magnetization data for $Eu_2ZnIrO_6$ a FM-like transition is observed at $T_N$ = 12 K (Fig. 5a). For a field of 1 T, field-cooled (FC) and zero-field cooled (ZFC) curves overlap over the whole temperature range from 2 K to 300 K. At lower field a splitting at low temperatures occurs. At applied fields of 0.1 T the splitting occurs at T≈8.5 K. At 0.01 T the splitting temperature increases to 12 K which is equal to $T_N$ (Fig. 5a). The M(H) data collected at T = 5 K show a hysteresis with a coercive field of 1460 Oe and a remanent magnetization of 0.33 $\mu_B$/f.u. (Fig. 5b). In comparison with $Gd_2ZnIrO_6$, the remanent field is significantly lower. This can be understood as an effect of the number of magnetic ions in the system. While $Ir^{4+}$ is the only magnetic contributor in $Eu_2ZnIrO_6$, there are two magnetic ions, $Gd^{3+}$ and $Ir^{4+}$, in $Gd_2ZnIrO_6$, contributing to the remanent magnetization. Since $Eu^{3+}$ is known to exhibit Van-Vleck paramagnetism [32], the Curie-Weiss law may not be applicable. Therefore, a Curie-Weiss analysis for $Eu_2ZnIrO_6$ is not shown.

### c) $Sm_2ZnIrO_6$

$Sm_2ZnIrO_6$ differs significantly from the two aforementioned compounds in that it adopts an antiferromagnetic order below $T_N$ = 13 K (Fig. 6a). The Néel temperature in this and all other AFM-materials in this work are defined as the position of the peak in the magnetization data. The antiferromagnetism remains stable even at high fields up to 7 T. This is also confirmed by the linear behavior in the M(H)-measurement at 5 K (Fig.6b). From the Curie-Weiss analysis, $\theta_{CW}$ = 6.7 ± 1.5 K and $\mu_{eff}$ = 1.7 ± 0.03 $\mu_B$/f.u. have been derived (Fig. 6c). The effective moment is close to the theoretical value, which is 2.11 $\mu_B$/f.u.

### d) $Nd_2ZnIrO_6$

This material shows the most complex and intriguing magnetic properties of all reported compounds. At low fields ($\mu_0 H \leq 0.1$ T) two transitions can be observed (Fig. 7c). An AFM-like kink at $T_1$ = 16.5 K is followed by a FM-like upturn starting at $T_2$ = 14.5 K for $\mu_0 H$ = 0.01 T. The AFM-like transition at $T_1$ = 16.5 K is not altered when the field is increased to $\mu_0 H$ = 0.1 T. $T_2$ is shifted to 11 K and instead of the upturn another downturn is observed. This downturn is particularly pronounced in the ZFC data. For 1 T ≤ $\mu_0 H$ ≤ 3 T only the AFM-like transition is found (Fig. 7b). The position of the AFM-peak shifts gradually from 16 K at 1 T to 13.5 K at 3 T. For fields higher than 3 T, the system is forced into a FM-like arrangement. The transition temperature is $T_C$ ≈ 14 K (Fig. 7a). In the field-dependent magnetization a corresponding spin-flop transition from a linear behavior at low fields to a curved shape at higher fields can be found at ≈ 3 T, when measured at 2 K (Fig. 7d). When the temperature is increased gradually to 10 K and 12.5 K, the curvature becomes less pronounced. At 15 K, a nearly linear M(H)-dependence is reached. The values derived from the Curie-Weiss fit of $Nd_2ZnIrO_6$ are $\theta_{CW}$ = -40.1 ± 1.3 K and $\mu_{eff}$ = 5.60 ± 0.03 $\mu_B$/f.u. (Fig. 7e). The negative Curie-Weiss temperature indicates fairly strong antiferromagnetic interactions. Its absolute value is higher than the ordering temperature in an external field of 1 T. The resulting frustration factor, calculated with the formula f = Θ$_{CW}$ /T$_N$, is 2.5. The theoretical value for μ$_{eff}$ is 5.40 μ$_B$/f.u., which is in close proximity to the experimental value.

**C. Specific heat**

The magnetic behavior of Nd$_2$ZnIrO$_6$ was further analysed by specific heat investigations as a function of temperature and applied magnetic field. The curves are shown in Fig. 8. In zero field a sharp magnetic transition occurs at T$_N$ = 16.5 K (determined from an entropy conserving construction), which is characteristic for a second order phase transition and which is gradually suppressed and broadened by the application of magnetic fields H ≥ H$_c$ (H$_c$ is the spin-flop field). The double transition at T$_{N1}$ = 16.5 K and T$_{N2}$ ~ 14.5 K observed in the temperature dependent magnetization curve at lowest fields is not discernible. Instead, an additional broad hump is observed at lower temperature T ~ 8 K. While the main transition at T$_{N1}$ = 16.5 K shifts to slightly lower temperatures as function of applied field in accordance with the dominantly antiferromagnetic ordering of this material, the hump seems to shift to slightly higher temperatures, pretending a merging of both features in *Cp*/T at higher fields. Further microscopic investigations are needed to fully understand the nature and field dependence of these two anomalies. Taking together the results from our magnetometry and bulk thermodynamic specific heat investigations, the second anomaly T < T$_{N1}$ suggests not be a second-order phase transition but more a crossover. Since the integral of the specific heat coefficient *Cp*/T yields the entropy of the investigated system, it seems as if the broad anomaly at lower temperature could designate (i) a small reordering of or between the two different magnetic sublattices (e.g. changes in the canting angle) or a freezing of either (ii) remaining spin dynamics of some (Ir-) magnetization components or of (iii) incoherent spin fluctuations of the ordered sublattices around their equilibrium positions. The latter scenario has been found via detailed macroscopic and microscopic studies on the mixed 3*d*-5*d* double perovskite material La$_2$CuIrO$_6$ [10]. Further investigations using microscopic techniques such as ESR and NMR will certainly be helpful to further understand the phase diagram of Nd$_2$ZnIrO$_6$.

**D. Neutron powder diffraction**

In order to unveil the magnetic structure of the ground state of Nd$_2$ZnIrO$_6$, neutron powder diffraction patterns at zero field were collected at three different temperatures, one in the paramagnetic region (30 K), one between the first and second magnetic ordering temperatures (15 K) and one below both magnetic transitions found in the magnetization measurements at low fields (4 K). The data at 30 K was used to determine the low temperature crystal structure of $Nd_2ZnIrO_6$ using Rietveld refinement (Fig. 9a and Fig. 10). When the temperature is lowered to 15 K, additional reflections emerge, most pronounced at 2θ = $25^0$, $44.5^0$ and $58.5^0$. These reflections grow in intensity when the temperature is further lowered to T = 4 K. Simultaneously, two additional new peaks become visible at $43^0$ and $57.5^0$ (Fig. 9b). The 4 K data was used to refine the magnetic ground state of $Nd_2ZnIrO_6$. The main magnetic Bragg peaks can be indexed as (1/2 1/2 1), (1/2 3/2 1) and (3/2 1/2 1). This indicates a magnetic propagation vector along (1/2 1/2 0). At first, the data was modeled assuming only the Nd-sublattice to order. Representational analysis combined with Rietveld refinement of the resulting magnetic structures, yielded a good fit for the irreducible representations labeled $\Gamma_1$ and $\Gamma_3$ in Kovalev's notation [34]. For either one of these irreducible representations an ordering of half of the $Nd^{3+}$-ions is observed and the ordered moment is estimated to be 3.4 $\mu_B$ for each $Nd^{3+}$-ion. This value is significantly higher than the ones found for $Nd^{3+}$ via NPD in similar compounds like $Nd_2NaRuO_6$ (2.3 $\mu_B$), $Nd_2NaOsO_6$ (1.6 $\mu_B$) [33] and $Nd_2O_3$ (1.9 $\mu_B$) [35]. The value of 3.4 µB for $Nd_2ZnIrO_6$ is close to the theoretical value (3 $\mu_B$) for $Nd^{3+}$. However, as illustrated by the literature-known compounds, this value is commonly not achieved in neutron diffraction studies. Instead, due to moment distribution to neighboring atoms through covalency effects as well as spin orbit coupling, lower values are often observed, as seen in the previous examples [33; 35]. Conclusively, it is unlikely that either $\Gamma_1$ or $\Gamma_3$ describe the ordering of the Nd-sublattice in $Nd_2ZnIrO_6$ correctly. Combining both irreducible representations into $\Gamma_1+\Gamma_3$ leads to a description of the magnetic ground state in which all $Nd^{3+}$-ions are ordered. They form a set of two distorted, interpenetrating cubic lattices with a G-type antiferromagnetic order (Fig. 11). In this model, all Nd-moments in $Nd_2ZnIrO_6$ lie within the ab-plane. The ordered moment of each Nd-ion in this model amounts to 2.4 $\mu_B$, which is reasonably close to the values found in the related compounds mentioned above. In a second step the magnetic moments of the $Ir^{4+}$-ions were included in the model, using the same irreducible representations as for the Nd-sublattice. The refined values of the coefficients representing the Ir-moments were negligibly low and the calculated diffraction pattern remained closely similar to the one excluding the Ir-magnetism (Fig. 9d and 9e). The quality of fit also remains almost unchanged for the models including (Rwp = 14.9) and excluding (Rwp = 15.1) the Ir-magnetism. From this result, and the fact that a good fit is achieved with the Nd-only model, it can be concluded that either the Ir-ions are unordered or the measurement is not sensitive enough to detect the comparably small moments of $Ir^{4+}$. Temperature-dependent intensity measurements were performed on the Bragg peaks corresponding to the order parameters 1/2 1/2 1 and 1/2 1/2 0 (Fig. 9f). In both cases, a steady increase of the intensity is found upon cooling the sample. The absence of kinks or abrupt changes of the slope indicate that there is only one magnetic phase present below 16 K in zero field, or the 2[nd] transition found in the magnetization measurements in low fields is not detectable by NPD, possibly because it involves the $Ir^{4+}$-moments.

## IV. DISCUSSION

As $Eu^{3+}$ is a non-magnetic ion, the magnetic properties of $Eu_2ZnIrO_6$ can be explained as pure Ir-sublattice magnetism. In comparison, the isostructural double perovskite $La_2ZnIrO_6$ shows a similar transition at slightly lower temperatures (7 K) [19]. In the La-compound, this behavior was found to be caused by a canted antiferromagnetic ground state [20]. A similar arrangement seems likely in $Eu_2ZnIrO_6$, since its net-ferromagnetic moment of approximately 0.4 $\mu_B$/f.u. in an external field of 1 T is close to the one found in $La_2ZnIrO_6$ [14; 19; 20]. The increased transition temperature for $A$ = $Eu^{3+}$ is explicable in terms of lattice effects. The lattice contraction caused by the smaller ionic radius of $Eu^{3+}$ compared to $La^{3+}$ leads to shorter distances between the magnetic $Ir^{4+}$-ions and may thereby strengthen the Ir-Ir-interactions. Hence the ordering of the Ir-sublattice occurs at a slightly higher temperature [29].

In $Gd_2ZnIrO_6$ on the other hand the rare earth ion $Gd^{3+}$ possesses an inherent magnetic moment. The observed ferromagnetic transition occurs at 24 K. This is the highest observed transition temperature in the $Ln_2ZnIrO_6$-series. $Gd_2O_3$, which has been found as a small impurity in this sample, orders antiferromagnetically at 1.6 K [38- 39], hence it can be excluded that the impurity phase is responsible for the transition found in the magnetization data. Since the magnetic properties of $Gd_2ZnIrO_6$ differ significantly from $La_2ZnIrO_6$ and $Eu_2ZnIrO_6$, which can hardly be explained solely by lattice effects, a direct influence of the magnetic moments of the $Gd^{3+}$-ions has to be assumed. Just like $Gd_2ZnIrO_6$, $Gd_2MgIrO_6$ has also been reported to adopt a ferromagnetic ground state [17]. Perovskites with a rare earth ion on the $A$-site as their only magnetic ion usually have a much lower transition temperature than $Gd_2ZnIrO_6$. For example, $GdAlO_3$ orders antiferromagnetically at 3.7 K [30]. Therefore, it is unlikely that the high transition temperature of 24 K in $Gd_2ZnIrO_6$ is caused solely by the presence of $Gd^{3+}$ on the $A$-site. Instead, an interplay of $Gd^{3+}$- and $Ir^{4+}$-magnetism is likely responsible for the magnetic transition.

$Sm_2ZnIrO_6$'s magnetic properties differ significantly from the aforementioned compounds. It adopts an antiferromagnetic order similar to $Sm_2MgIrO_6$ [17]. The similarity of $Sm_2ZnIrO_6$ and $Sm_2MgIrO_6$ points towards the $Sm^{3+}$-sublattice being the driving force for the AFM order. Since the ordered moment in $Sm_2ZnIrO_6$ is significantly smaller than the one of the pure Ir-sublattice in $Eu_2ZnIrO_6$ and $La_2ZnIrO_6$ [14; 19; 20], the canted antiferromagnetic order of the $Ir^{4+}$ seems to be suppressed or masked by Sm-Ir interactions.

The compound with the largest *A*-cation, $Nd_2ZnIrO_6$, shows the most intriguing magnetic properties. The magnetic ground state of this material in zero-field has been analyzed by NPD. It was found to be a complex ordering of the Nd-sublattice (Fig. 11) with little or no contribution from the Ir-magnetism. However, magnetization data show that the magnetic phase diagram of $Nd_2ZnIrO_6$ is more diverse. In fact, a strong field dependence is observed, including a clearly observable metamagnetic transition in the M(H) data (Fig. 7d). While the Ir-magnetism may be negligible in zero-field according to NPD, it seems likely that contributions of the $Ir^{4+}$- sublattice play a role in higher fields, leading to this complex magnetic behavior. The magnetic properties of $Nd_2ZnIrO_6$ are strikingly similar to $Nd_2NaOsO_6$ [15]. In the osmate, a simultaneous order, in which the Ir-spins adopt a type-I antiferromagnetic order while the Nd moments adopt a canted antiferromagnetic arrangement, was proven by neutron scattering experiments [33]. The similarity of both materials points towards a general feature of $Nd^{3+}$, which experiences more complex interactions with 5*d*-transition metal ions than the other lanthanides, leading to complex magnetic behavior in both the iridate and the osmate double perovskite.

## V.     CONCLUSIONS

We have successfully synthesized the double perovskite series $Ln_2ZnIrO_6$ (Ln = Nd, Sm, Eu & Gd) and investigated the magnetic ground states for the different lanthanide ions. These materials show a variety of effects arising from the interplay of the 4*f* -ions on the *A*-site with the strongly spin-orbit coupled $Ir^{4+}$-ions on one *B*-site. $Gd_2ZnIrO_6$ and $Eu_2ZnIrO_6$ have ferromagnetic or canted antiferromagnetic ground states. $Sm_2ZnIrO_6$ shows antiferromagnetic order, likely arising from Sm-Ir interactions. Finally, $Nd_2ZnIrO_6$ shows particularly strong correlation between the two magnetic sublattices, leading to a strongly field dependent ground state. Neutron powder diffraction measurements confirms the presence of long-range magnetic order. Specific heat investigations for $Nd_2ZnIrO_6$ show a sharp magnetic transition at $T_N$ = 16.5 K, which is gradually suppressed and broadened by the application of magnetic fields.

## VI.     ACKNOWLEDGEMENTS

This work was supported by the Deutsche Forschungsgemeinschaft (DFG) through project B01 of the SFB 1143 (project-id 247310070), Grant No. AS 523/4-1 and Grant No. WU 595/3-3. A portion of this research used resources at the High Flux Isotope Reactor, a DOE Office of Science User Facility operated by Oak Ridge National Laboratory.

**Captions**

**FIG. 1:** X-ray diffraction pattern of $Ln_2ZnIrO_6$ with Ln = Gd, Eu, Sm, Nd. For $Gd_2ZnIrO_6$, the impurity peak of $Gd_2O_3$ is marked.

**FIG. 2**: Evolution of cell parameters a, b and c. The lines serve as guide to the eye.

**FIG. 3**: Topography (SE mode; left) and chemical contrast (BSE mode; right) images of $Eu_2ZnIrO_6$ taken by SEM. A comparison of both pictures show that color differences in the chemical contrast image stem from the uneven surface. No chemical inhomogeneity is evident.

**FIG. 4**: Magnetization data of $Gd_2ZnIrO_6$. a) Field-cooled (FC) and zero-field cooled (ZFC) temperature-dependent magnetization in external fields of 0.1, 1 & 3 T. b) Field-dependent magnetization at 5 K. The inset shows the full range up to 5.5 T. c) Curie-Weiss analysis in the temperature range 150 K - 300 K.

**FIG. 5**: Magnetization data of $Eu_2ZnIrO_6$. a) Field-cooled (FC) and zero-field cooled (ZFC) temperature-dependent magnetization in external fields of 0.01, 0.1 & 1 T. A splitting of the FC and ZFC curves occurs at low fields. b) Field-dependent magnetization at 2 K and 11 K.

**FIG. 6**: Magnetization data of $Sm_2ZnIrO_6$ a) Field-cooled (FC) and zero-field cooled (ZFC) temperature dependent magnetization. An antiferromagnetic transition is observable at $T_N$ = 13 K. b) Field dependent magnetization at 5 K. c) Curie-Weiss analysis in the temperature range 150 K - 300 K.

**FIG. 7**: Magnetization data of $Nd_2ZnIrO_6$. a)-c) Field-cooled (FC) and zero-field cooled (ZFC) magnetizations of $Nd_2ZnIrO_6$ at different external fields. a) At high fields a ferromagnetic-like transition is found at T ≈ 14 K. b) At intermediate fields antiferromagnetic behavior is observed. The transition temperature is field-dependent and shifts from 16 K at 1 T to 13.5 K at 3 T. c) At low fields two transitions are observed. d) Field dependent magnetization of $Nd_2ZnIrO_6$. At 2 K (inset) a metamagnetic transition is observed at $\mu_0H$ ≈ 3 T. e) Curie-Weiss analysis in the temperature range 150 K - 300 K.

**FIG. 8**: Temperature dependence of the specific heat capacity of $Nd_2ZnIrO_6$ for different applied magnetic fields up to 9 T. In the inset the specific heat coefficient $C_p$/T is shown for temperatures around the phase transition. Applied magnetic fields slightly suppress the transition; for details see text.

**FIG. 9**: Powder neutron diffraction data of $Nd_2ZnIrO_6$. a) Rietveld refinement of the data set collected at 30 K. The atomic structure was derived from this data set (Fig. 10, table 2). b) Comparison of the data sets collected at 30 K, 15 K and 4 K. A set of magnetic Bragg-peaks arises when the temperatures is lowered. c) and d) Rietveld refinements of the data sets collected at 15 K and 4 K respectively. The magnetic contribution of the $Nd^{3+}$-cations is included in the model. For both temperatures the same magnetic Bragg-peaks are found. Their intensities are higher at 4 K. e) Rietveld refinement of the data set collected at 4 K. The model includes contributions of $Nd^{3+}$- and $Ir^{4+}$-magnetism. No significant differences to the refinement excluding the $Ir^{4+}$-magnetism (d) are observed. f) Temperature dependence of the Bragg-peaks corresponding to order parameters ½ ½ 1 and ½ ½ 0 (inset). A steady increase is observed upon decreasing the temperature, indicating only one observable magnetic transition.

**FIG. 10**: Crystal structure of $Nd_2ZnIrO_6$ at 30 K, derived from powder neutron diffraction data.

**FIG. 11**: Magnetic structure of $Nd_2ZnIrO_6$. a) the double perovskite structure with the magnetic moments of the Nd-ions shown. b) the magnetic Nd-sublattice. Blue and orange atoms correspond to the two interpenetrating AFM sublattices.

**Table. I**: EDXS results for all members of the $Ln_2ZnIrO_6$ series. Oxygen contents are excluded due to their high inaccuracy when measured by EDXS.

**Table. II**: Structural parameters of $Nd_2ZnIrO_6$ derived from the NPD data collected at 30 K with λ = 1.54 Å.

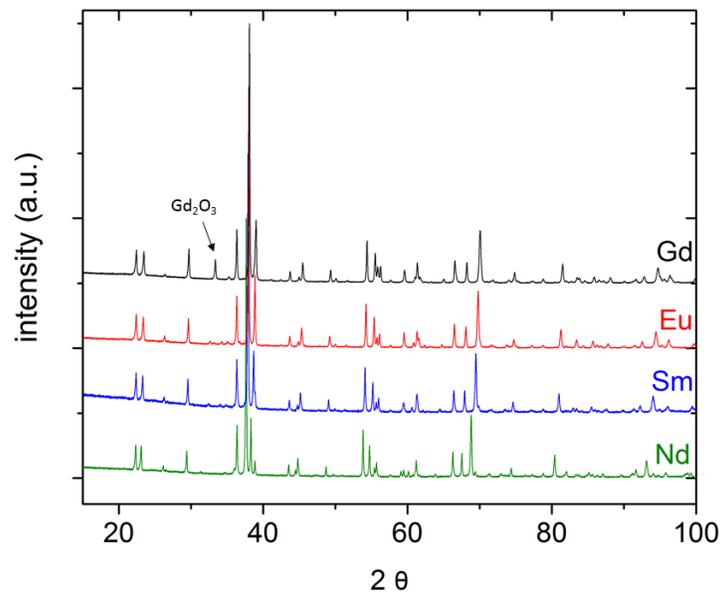

**Fig. 1.**

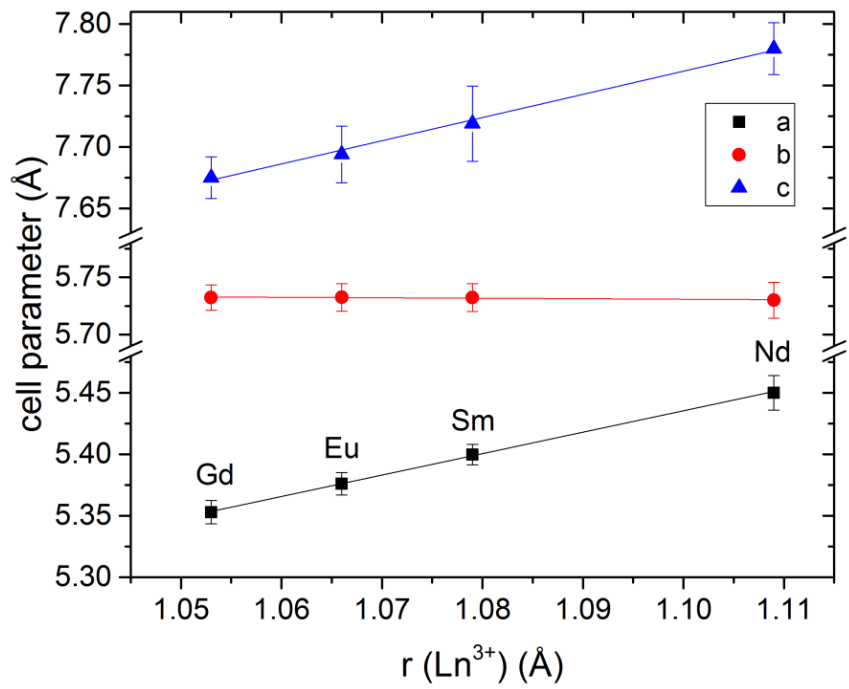

**Fig. 2.**

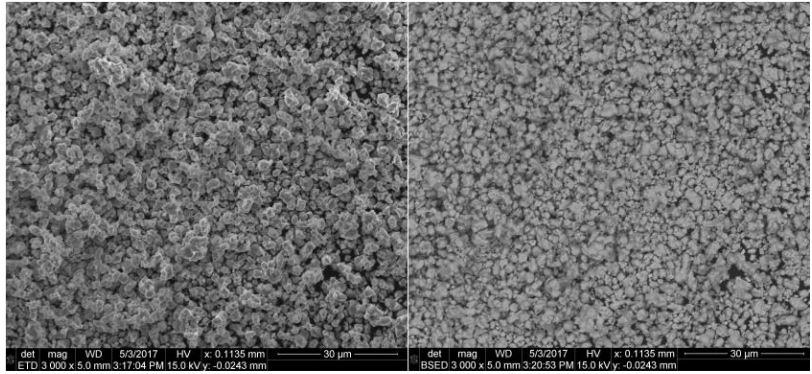

**Fig. 3.**

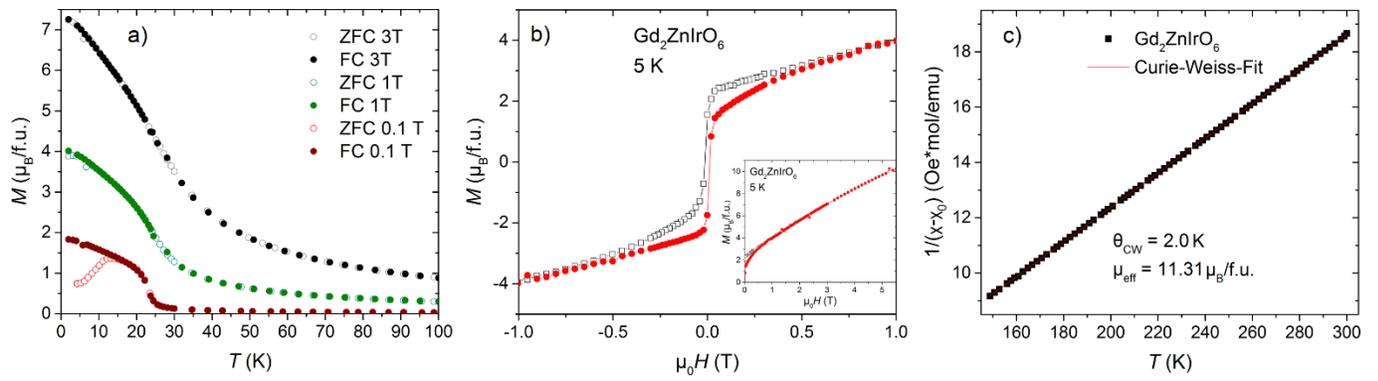

**Fig. 4.**

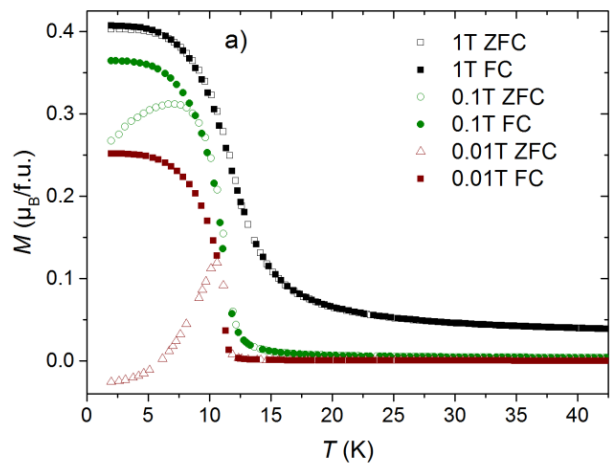

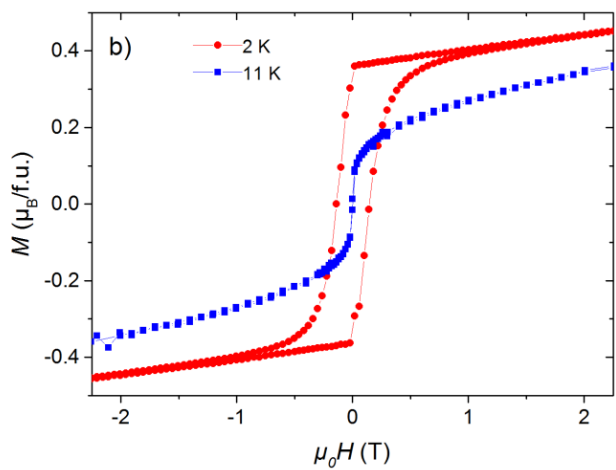

**Fig. 5.**

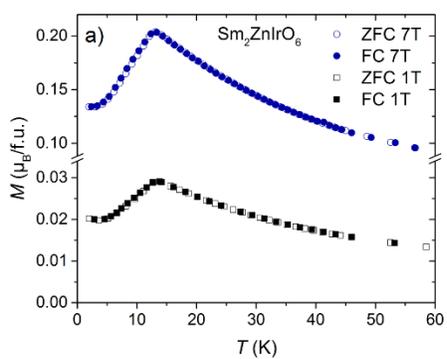 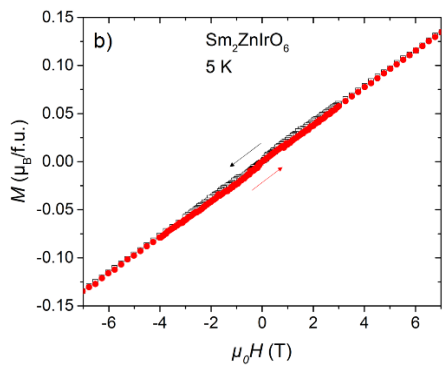 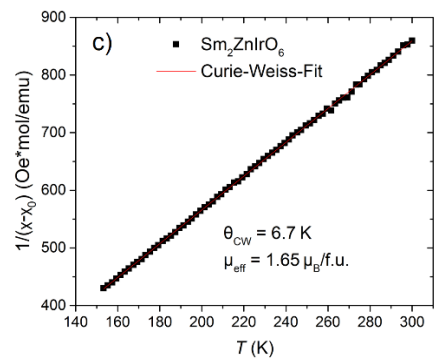

**Fig. 6.**

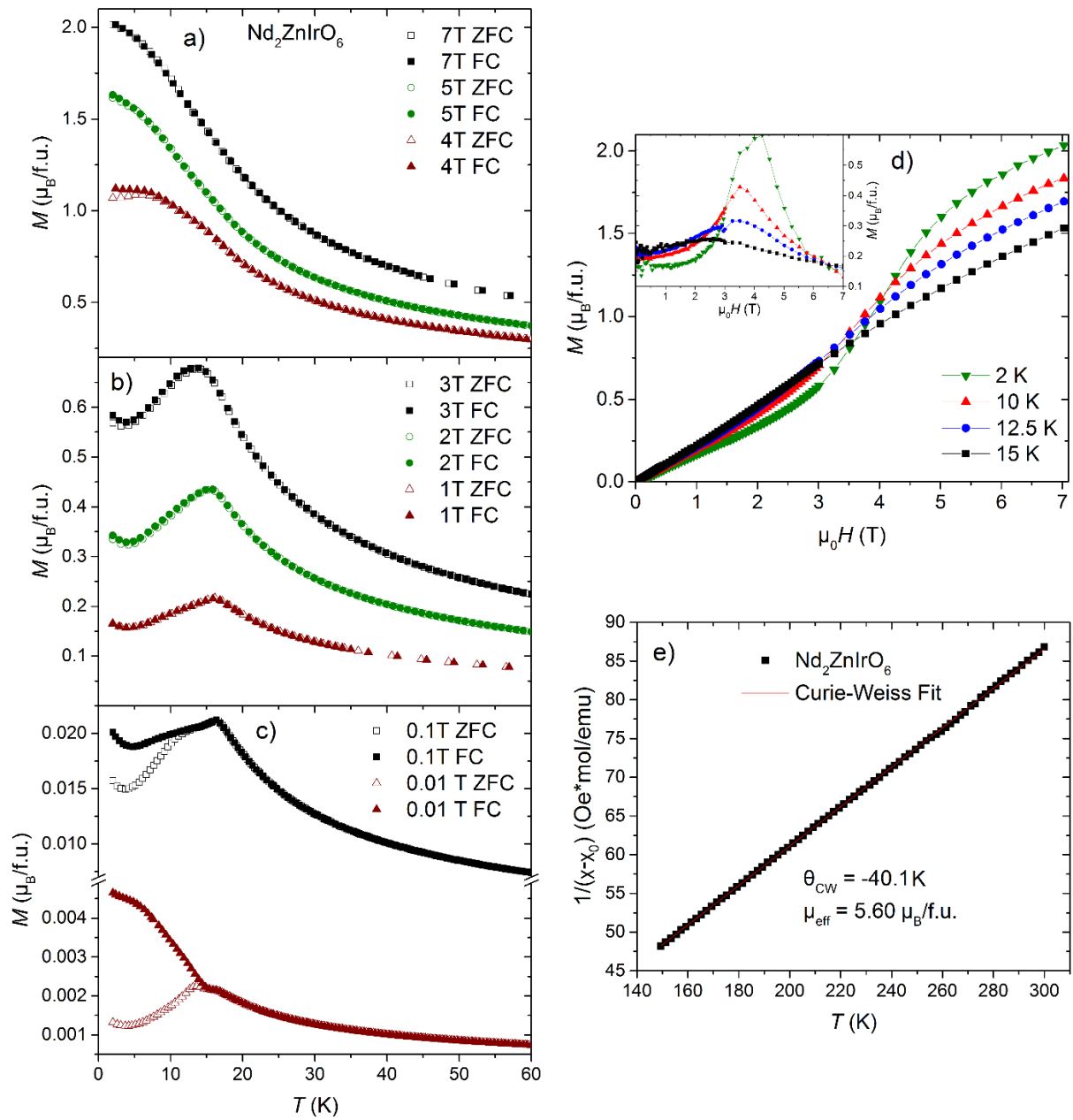

**Fig. 7.**

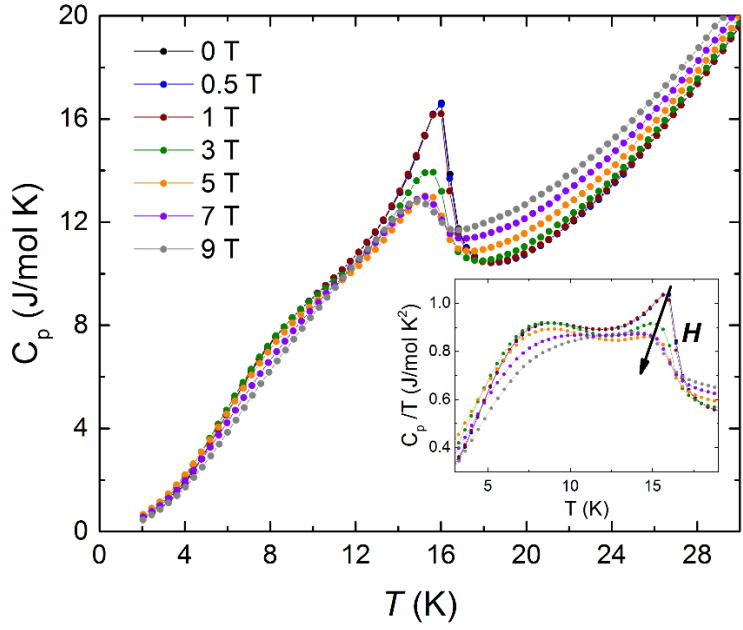

**Fig. 8.**

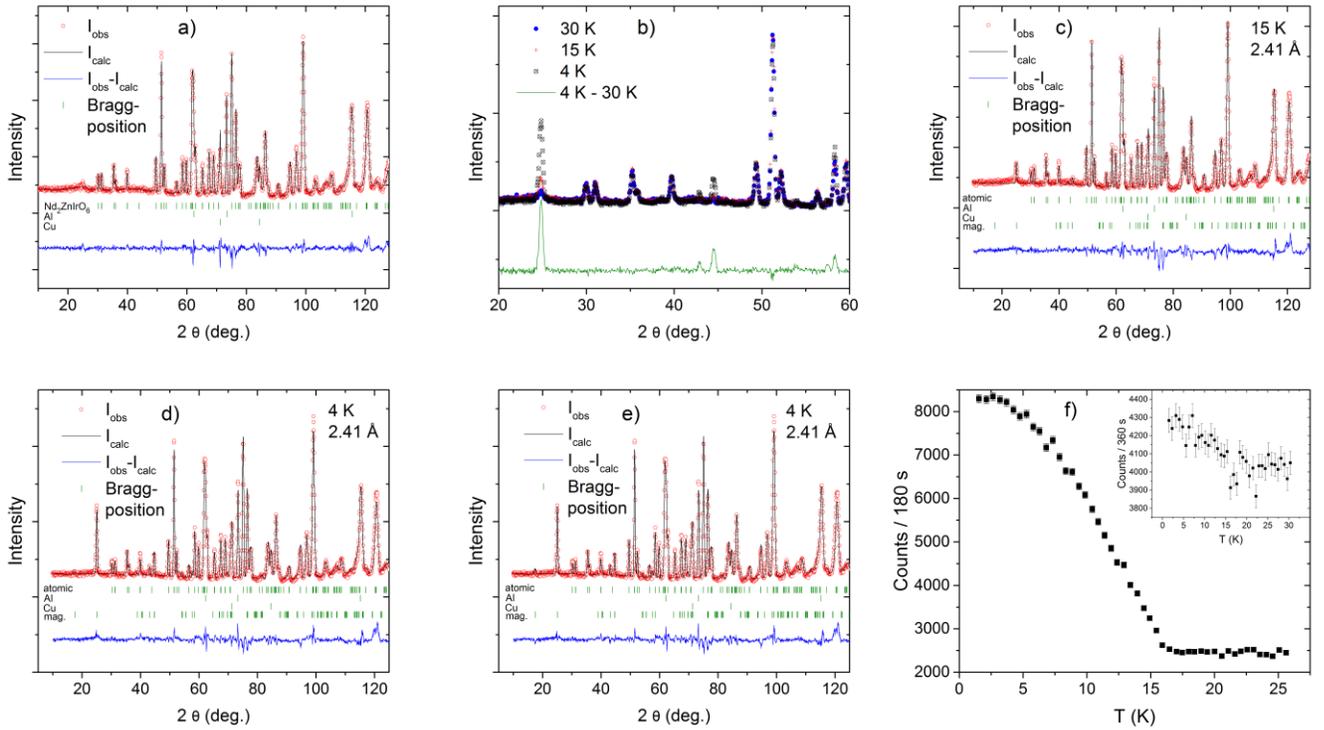

**Fig. 9.**

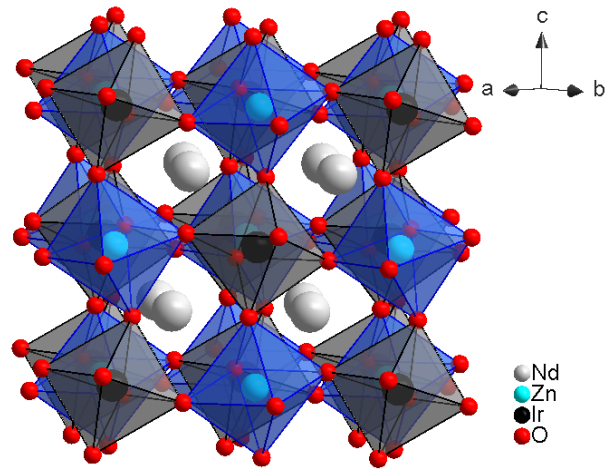

**Fig. 10.**

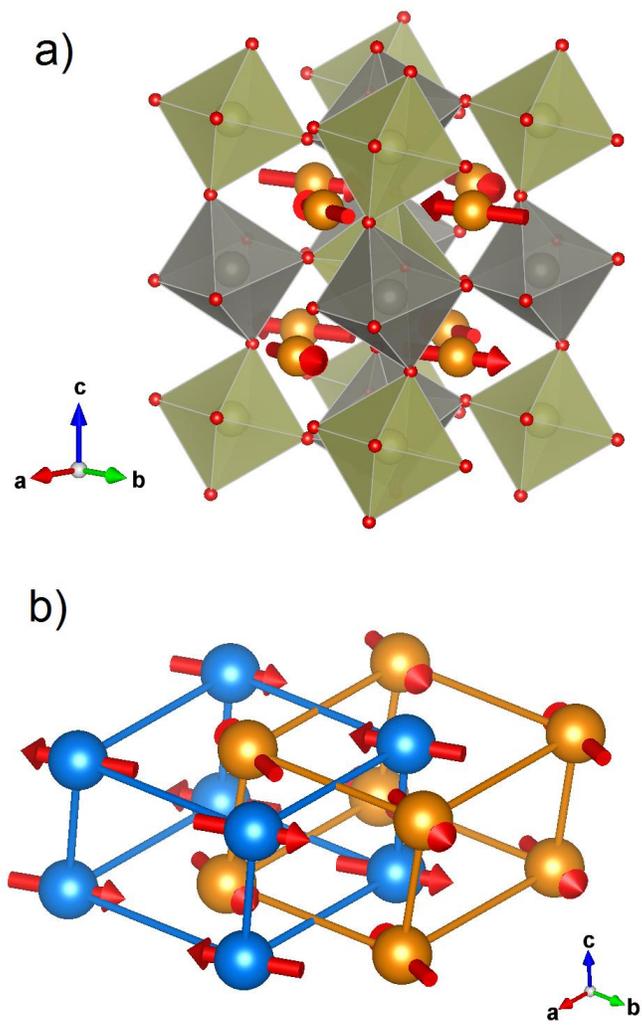

**Fig. 11.**

| Ln | at%(Ln) | at%(Zn) | at%(Ir) |
|---|---|---|---|
| Nd | 50.9±1.3 | 25.1 ± 1.3 | 24.1 ± 1.2 |
| Sm | 50.7 ± 1.8 | 24.7 ± 1.1 | 24.6 ± 1.5 |
| Eu | 51.6 ± 1.1 | 24.2 ± 1.8 | 24.2 ± 1.6 |
| Gd | 51.7 ± 1.7 | 24.2 ± 1.6 | 24.1 ± 1.7 |
| nominal | 50 | 25 | 25 |

Table I.

| | space group | $P2_1/n$ |
|---|---|---|
| | a | 5.4586(1) Å |
| | b | 5.7463(2) Å |
| | c | 7.7892(2) Å |
| | $\beta$ | 89.939(8) |
| | $\chi^2$ | 1.07 |
| | $R_{wp}$ | 12.9 % |

| atom | site | a | b | c |
|---|---|---|---|---|
| Nd | 4e | 0.0194(8) | 0.9383(5) | 0.249(5) |
| Zn | 2d | 0.5 | 0 | 0 |
| Ir | 2c | 0 | 0.5 | 0 |
| O1 | 4e | 0.187(1) | 0.199(2) | 0.062(1) |
| O2 | 4e | 0.195(2) | 0.198(2) | 0.460(1) |
| O3 | 4e | 0.6010(9) | 0.0316(8) | 0.263(1) |

Table II.